# Low Resistance GaN/InGaN/GaN Tunnel Junctions

Sriram Krishnamoorthy, [1,a)] Fatih Akyol[1], Pil Sung Park[1] and Siddharth Rajan[1,2, a)]

[1]Department of Electrical & Computer Engineering, The Ohio State University, Columbus, OH 43210
[2]Department of Materials Science and Engineering, The Ohio State University, Columbus, OH 43210

**Abstract**: Enhanced interband tunnel injection of holes into a p-n junction is demonstrated using p-GaN/InGaN/n-GaN tunnel junctions with a specific resistivity of $1.2 \times 10^{-4}$ $\Omega$ cm$^2$. The design methodology and low-temperature characteristic of these tunnel junctions is discussed, and insertion into a p-n junction device is described. Applications of tunnel junctions in III-nitride optoelectronics devices are explained using energy band diagrams. The lower band gap and polarization fields reduce tunneling barrier, eliminating the need for ohmic contacts to p-type GaN. This demonstration of efficient tunnel injection of carriers in III-nitrides can lead to a replacement of existing resistive p-type contact material in light emitters with tunneling contact layers, requiring very little metal footprint on the surface, resulting in enhanced light extraction from top emitting emitters.

a) Authors to whom correspondence should be addressed.
Electronic mail: krishnamoorthy.13@osu.edu, rajan@ece.osu.edu

This paper describes low-resistance III-nitride tunnel junctions using polarization-engineered p-GaN/InGaN/n-GaN junctions for applications in III-nitride devices. While interband quantum mechanical tunneling is exploited in different material systems to enhance device performance or functionality,[1-3] applications in the III-nitride devices have been relatively few[4,5]. In the past, the large band gap of III-nitrides (specifically Ga(Al)N) resulted in low tunneling probability and high tunneling resistance in interband tunnel junctions, with additional voltage drop that were unacceptably high for application in devices. However, polarization engineering can surmount the intrinsic barriers to tunneling in these materials. This is achieved by aligning valence and conduction bands on either side of the junction by utilizing the high electric fields from the interfacial polarization-induced dipoles at GaN/AlN/GaN,[6,7] or GaN/InGaN/GaN[8-10] junctions. The resulting reduction in depletion width (or tunneling distance) as well as the reduction in energy barrier when using a lower band gap InGaN barrier leads to a high reverse current density in GaN tunnel junctions,[8] and forward tunneling characteristics with negative differential resistance in III- nitride tunnel diodes[9,10].

Tunnel injection of holes in III- nitride devices could enable a number of electronic and optoelectronic device designs that are otherwise limited due to resistive p-type materials and p-contacts. Fig. 1 illustrates two such applications of tunnel junctions in III- nitride devices. Efficient interband tunnel junctions act as a carrier conversion center enabling tunnel injection of holes into the active device layer. In a reverse biased tunnel junction, an electron in the valence band of p-type material tunnels into the empty states in the conduction band of the n-type material, leaving behind a hole in the p-type material, leading to hole injection into the p-type material. Fig 1(a) shows a top emitting LED structure with a tunnel junction to enable a n-type top contact layer. This device structure enables device designers to replace the relatively more challenging p-contact with low resistance n-type, which could be very important for larger band gap AlGaN. In the case of p-type top contacts, full metal coverage is required to make contact to the semiconductor since p-type GaN has high sheet resistance (low mobility and carrier concentration). With tunnel junctions, where the top contact layer is n-type GaN with low sheet resistance and excellent current spreading, smaller metal footprint on the top surface would be sufficient, thus enhancing light extraction efficiency. In case of deep UV LEDs,[11] tunnel junctions can eliminate the need for thick p-GaN which is otherwise required for hole injection into p-AlGaN. Since the low band gap layer required is thinner than the thick p-GaN cap, the associated absorption loss will also be reduced. The non-equilibrium tunnel-injected holes could also be used to substitute thermally ionized holes completely, enabling bipolar devices without any p-type doping [12].

Efficient tunnel junctions could also help realize new device topologies. Emitters with reversed polarization are expected to improve carrier injection and confinement in the quantum well LEDs[13-15] and solar cells[16]. The p-down structure necessary to realize these reverse polarization structures along the +c orientation, has until now been challenging due to current crowding and poor ohmic p-contact on etched surfaces. A tunnel junction based structure (Figure 1(b)) would overcome the both p-region spreading and contact issues and enable inverted polarity LEDs for potentially higher efficiency. Tunnel junctions are necessary for connecting multiple active regions in series (Figure 1 (c)), such as in multi-color LEDs,[17] and multi-junction solar cells[18]. This is especially attractive in the III- nitride material system with a large range of band gaps accessible, enabling monolithic integration for applications such as white LEDs.

The equilibrium band diagram of a p-GaN/InGaN/n-GaN tunnel junction is shown in Figure 2 (a). The tunneling barriers between the undepleted p-GaN and n-GaN regions consist of the depletion regions in the GaN on the n-side, the InGaN quantum well itself, and the depletion on the p–side and it is necessary to take each of these three barriers into account. We first consider the InGaN quantum well. For a Ga-face p-down or N-face p-up structure, the polarization field due to the sheet charges at GaN/InGaN interface and the depletion field of the heavily doped p-n junction are in the same direction, and this configuration is favorable for tunneling. The "critical" thickness ($t_{cr}$) of InGaN barrier (Figure 2 (b)) required for the band bending estimated from the polarization sheet charge at GaN/InGaN interface was found to be less than 10 nm even for InN molefraction as low as 18%[8]. As the indium composition is increased, the polarization sheet charge at GaN/InGaN interface increases and the thickness of InGaN barrier necessary to drop the p-n junction built-in potential is reduced. The lower depletion layer thickness (and tunneling distance) and bandgap increase the tunneling probability across the InGaN layer.

However, band discontinuities at the GaN/ InGaN heterojunction leads to additional depletion in GaN which is expected to reduce the overall tunneling probability (inset of Fig. 2a). We estimate the tunneling probability (T) due to the three barriers by multiplying the individual probabilities calculated using a Wentzel-Kramers-Brillouin (WKB) approximation. We note here that the semi-classical approach may have limitations since wavefunctions in the three regions are coupled and potential variation is sharp but use it as a guide to device design. The probability due the intraband tunneling in n-type ($T_n$) and p-type GaN ($T_p$) are related to conduction band ($\Delta E_c$) and valence band discontinuity ($\Delta E_v$) respectively, and therefore decrease as the indium composition is increased. The interband tunneling through the InGaN barrier ($T_{InGaN}$) increases with indium composition, as discussed earlier. For interband tunneling probability estimation, the Kane model was used to calculate the wave vector within the barrier[19]. The tunneling probability estimation assumes negligible reverse bias across the tunnel junction required for tunneling that does not change the band diagram significantly from the equilibrium.

The overall tunneling probability ($T_{net}$) of the device structure for a given InN molefraction, is the product of three tunneling probabilities estimated as

$$T_{InGaN} = \exp\left\{-2\int_0^{t_{cr}} \left[\sqrt{\frac{2m^*_{InGaN}\left(\left(\frac{E_{g,InGaN}}{2}\right)^2 - \left(\frac{E_{g,InGaN}}{2} - \frac{E_{g,InGaN}t}{t_{cr}}\right)^2\right)}{\hbar^2 E_{g,InGaN}}}\right]dt\right\} \quad (1),$$

$$T_n = \exp\left\{-2\int_0^{x_n}\sqrt{\frac{m_e^* q^2 N_D t^2}{\hbar^2 \varepsilon}}dt\right\} \quad (2),$$

$$T_p = \exp\left\{-2\int_0^{x_p}\sqrt{\frac{m_h^* q^2 N_A t^2}{\hbar^2 \varepsilon}}dt\right\} \quad (3),$$

is given by

$$T_{net} = T_{InGaN} \, T_n \, T_p \tag{4}$$

Here, $x_n$ and $x_p$ refer to the depletion region width in n-GaN and p-GaN respectively, given by,

$$x_n = \sqrt{\frac{2\varepsilon \Delta E_c}{q^2 N_D}} \tag{5},$$

$$x_p = \sqrt{\frac{2\varepsilon \Delta E_V}{q^2 N_A}} \tag{6}.$$

$N_A$ and $N_D$ refer to the acceptor and donor doping density, $m^*_e$ ($m^*_h$) refers to the effective mass of electron (hole), and $\varepsilon$ is the permittivity.

These calculations provide some guidelines for the design of GaN/InGaN/GaN tunnel junctions. The result of this calculation is shown in Figure 2 (b), which shows that for the doping density assumed ($N_A$ = 1 X 10$^{19}$ cm$^{-3}$, $N_D$ = 5 X 10$^{19}$ cm$^{-3}$), indium composition of approximately 25-30% results in the highest tunneling probability. Our calculation shows that there is an *optimal* composition that depends upon doping density – higher doping density would enable a higher composition and higher probability. In addition, since the conduction band discontinuity is higher than the valence band discontinuity,[20] increasing n-type doping density would have a much more important effect on tunneling probability. The doping density is limited by considerations of surface morphology, which can degrade at high doping density, as well as intrinsic solubility.

To demonstrate the performance of polarization-engineered GaN/InGaN/GaN tunnel junctions in a real device, the structure shown in Figure 3(a) was grown. The structure consists of a GaN p-n junction on top of a p-GaN/4 nm In$_{0.25}$Ga$_{0.75}$N/n-GaN tunnel junction, so that tunneling is used to inject holes into the p-type layer of the p-n junction. The samples were grown[21,22] by N$_2$-plasma assisted molecular beam epitaxy (Veeco Gen930 system) on N-face free standing GaN templates[23]. The polarity of the surface was confirmed post-growth by 3 X 3 surface reconstructions observed in-situ using reflection high energy electron diffraction (not shown here). Ti (20 nm) / Au (200 nm) ohmic contacts were deposited for the top n-contact, followed by mesa isolation, and then evaporation of the bottom n-type contact. No p-type contact formation is necessary since the holes are injected by the tunnel junction.

Electrical characteristics of this p-contact less p-n junction device (50 μm X 50 μm) are shown in Fig. 3(b), and show p-n junction behavior with rectification. The total series resistance of the device in forward bias was found by fitting the linear region of the forward bias characteristics to be 4.7 X $10^{-4}$ $\Omega cm^2$. This total resistance of the device is the sum of the contact resistance to n-type GaN, series resistance in the p-GaN and n-GaN regions, and the resistance of the tunnel junction. The contact resistance of the top n-type contact was estimated to be 3.5 X $10^{-4}$ $\Omega cm^2$ using transfer length measurement patterns on the n-type region. The specific resistivity of the tunnel junction is therefore lower than $1.2 \times 10^{-4}$ $\Omega cm^2$, which is the lowest observed resistance for the III-nitride tunnel junction. At a forward current density of 100 A/cm$^2$, the voltage drop across the p-n junction is 3.05 V, and the voltage drop across the TJ is 12 mV.

To further investigate tunnel injection of holes, low temperature IV measurements were performed on the p-n junction sample, and the results are summarized in Fig. 3(c). As expected, the reverse leakage of the p-n junction reduces on lowering the temperature as it can be clearly seen in the log I- V plot shown in the inset of Fig. 3(c). Under forward bias, low series resistance was observed even at low temperatures where hole freeze-out is expected.

Reported values for tunnel junction resistivity in various material systems as a function of the material band gap is shown in Figure 4[7, 24-32]. As the band gap of the material increases, the tunneling resistivity obtained increases exponentially, as expected from semi-classical WKB theory. However, with polarization engineering demonstrated in this work, the tunneling resistivity for GaN can overcome the band gap limits, bringing the resistance down to ~ $10^{-4}$ $\Omega cm^2$. Although, this device demonstration was performed on N-face orientation with the p-type layer down structure, similar characteristics would be obtained from a Ga-face n-GaN/InGaN tunnel junction on a +c-plane oriented p-up p-n junction or LED structure. This implies that the low resistance tunnel junctions demonstrated in this work can be directly extended to commercial c-plane LEDs. The low resistance n-type layer on top would enable several of the applications described earlier, and the reduced metal footprint on the n-type GaN could greatly enhance light extraction from top and eliminate the need for flip-chip processing[4, 5].

Polarization-engineered tunneling demonstrated here could enable tunnel-injected holes or electrons in other wide-bandgap material systems where high density of spontaneous and piezoelectric interfacial polarization charges are available, such as the (Mg,Zn)O system. Such an approach could overcome the fundamental doping limitations that have until now limited the device performance for such material systems.

In summary, low resistance ($1.2 \times 10^{-4}$ Ω cm$^2$) GaN/ InGaN based tunnel junctions were demonstrated in a GaN p-n junction. The low tunneling resistance obtained in this work demonstrates the promise of tunnel junctions for c-plane visible and UV LEDs. We have demonstrated that polarization engineered tunnel junctions can overcome p-type doping issues and enable increased functionality and performance for both electronic and optoelectronics III- nitride devices.

**Acknowledgement:**

The authors would like to acknowledge Prof. Emre Gur and Prof. Steven A Ringel for low temperature measurements shown in this work. We would like to acknowledge funding from Office of Naval Research under the DATE MURI program (Program manager: Paul Maki), and the National Science Foundation (DMR-1106177).

**Figure captions**:

**Figure 1**: (Color online) (a) Band diagram and device structure showing tunnel injection of holes in a III- nitride LED. (b) Reverse polarization structure enabled by tunnel junction (c) band diagram and device structure showing cascaded active regions with tunnel junction interconnects.

**Figure 2**: (Color online) (a) Equilibrium band diagram of a polarization engineered GaN/InGaN/GaN TJ; **Inset**: Tunneling barrier due to band discontinuities $\Delta E_c$ and $\Delta E_v$ (b) Tunneling probability estimation for the three different tunneling components, namely the inter-band tunneling through InGaN barrier, intraband tunneling through depletion regions in GaN ($N_A = 1 \times 10^{19}$ cm$^{-3}$, $N_D = 5 \times 10^{19}$ cm$^{-3}$). Overall tunnel probability is higher for InN molefraction in the range of 25%- 40%. Thickness of InGaN barrier ($t_{cr}$) as a function of InN molefraction is plotted in the right hand ordinate axis.

**Figure 3**: (Color online) (a) Epitaxial stack of a GaN p-n junction with a GaN/InGaN/GaN tunneling contact layer to p-GaN. Circuit model showing two back to back connected diodes and the various resistance components. (b) Electrical characteristics of the p-contact less p-n junction device showing negligible additional voltage drop across the tunnel junction. **Inset:** Resistance measurement across TLM pads with different spacing, used to extract specific top contact resistivity. (c) Temperature dependent I-V characteristics of the device showing efficient hole tunnel injection even at low temperatures. **Inset**: Temperature dependent Log I-V characteristics of the device.

**Figure 4**: (Color online) Tunnel junction resistivity in different material systems. In this work, the lowest tunneling resistivity for GaN is reported.

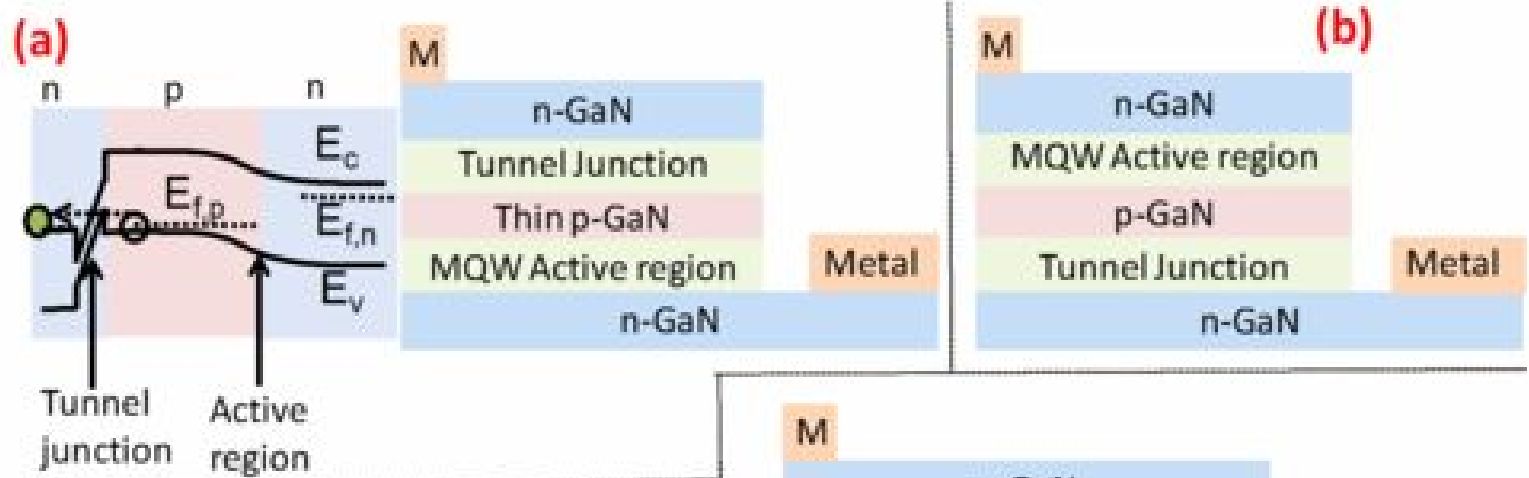
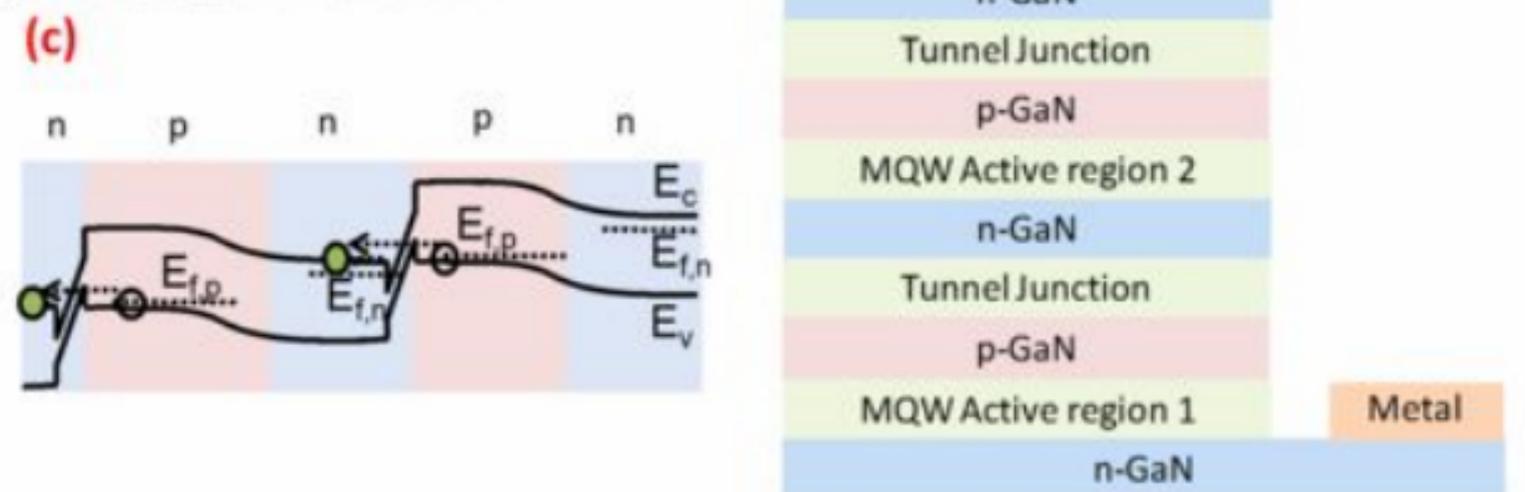

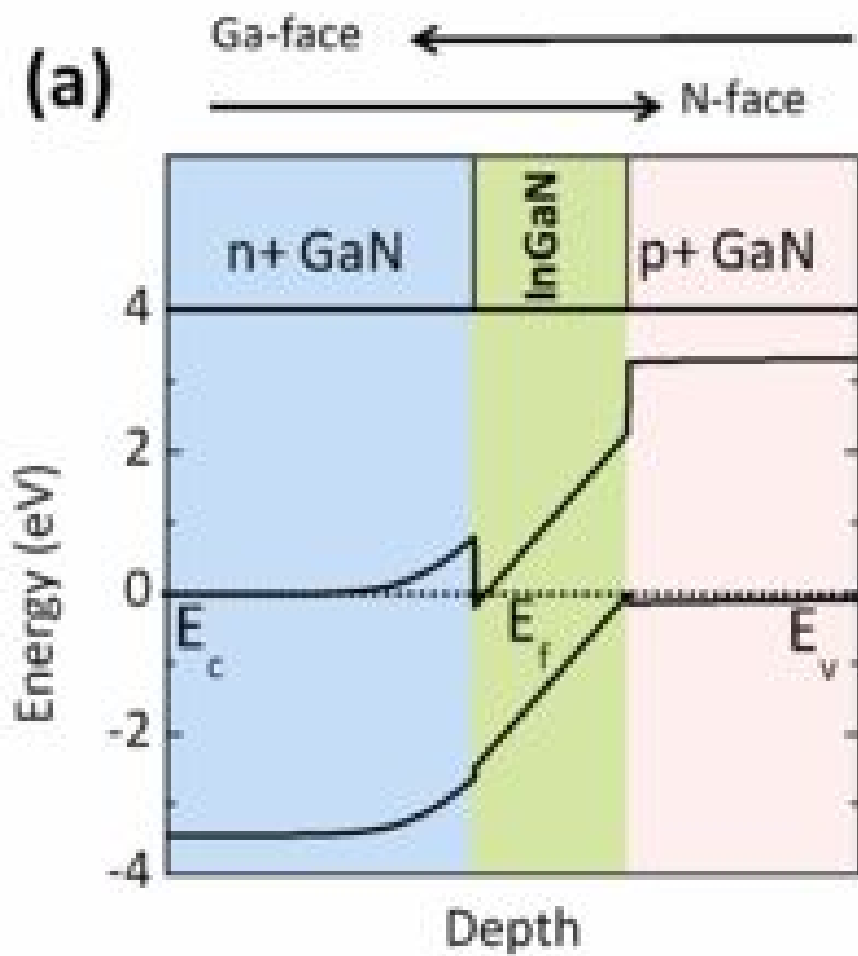
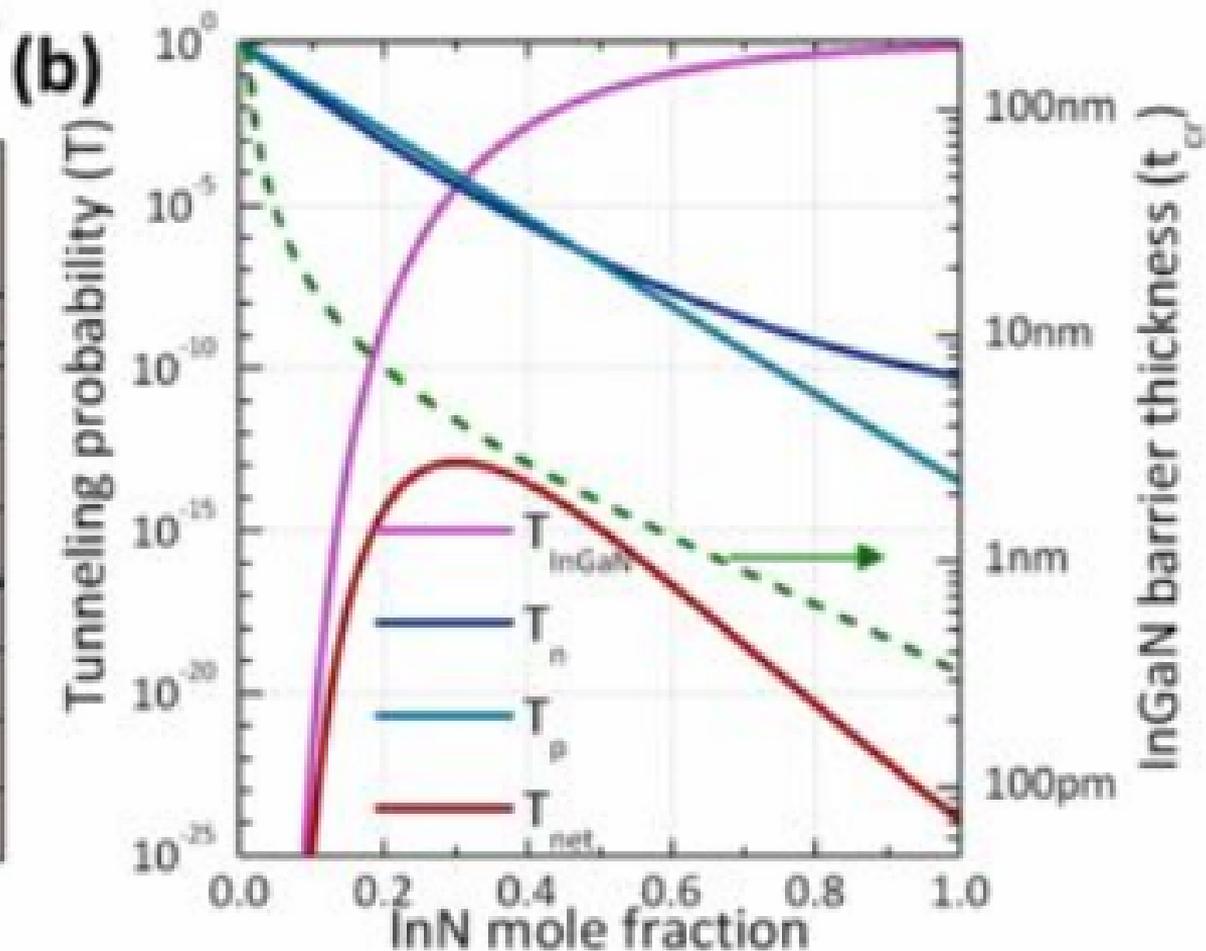

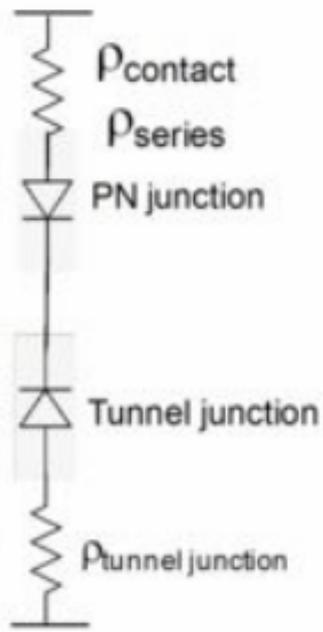
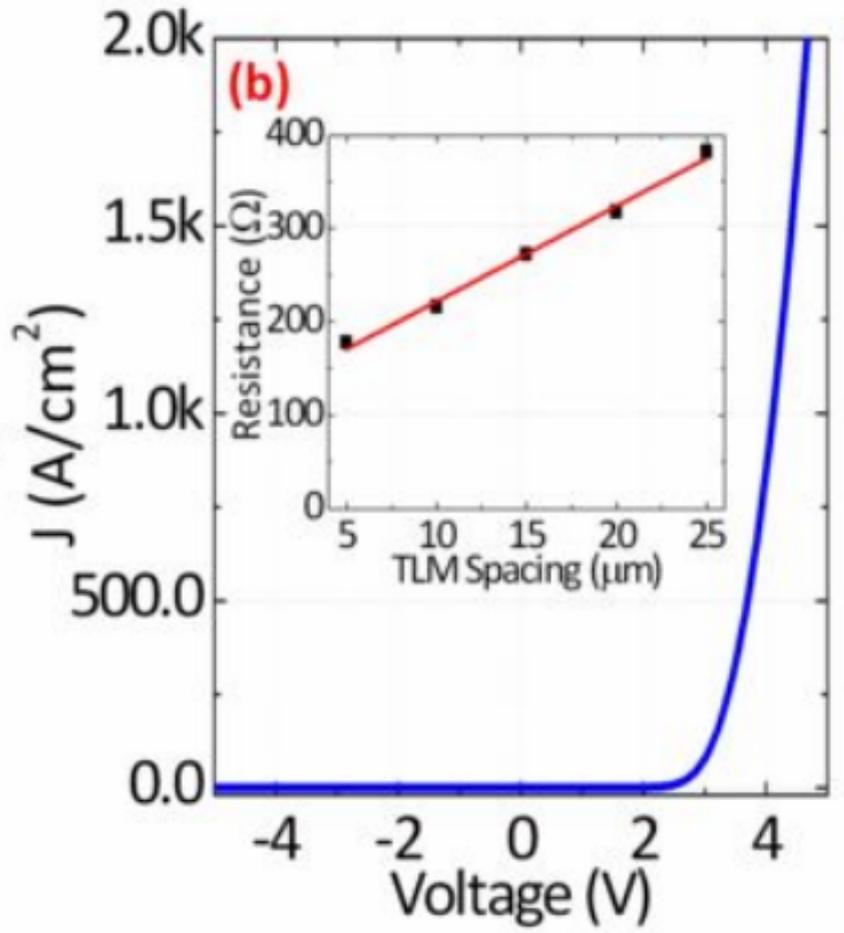
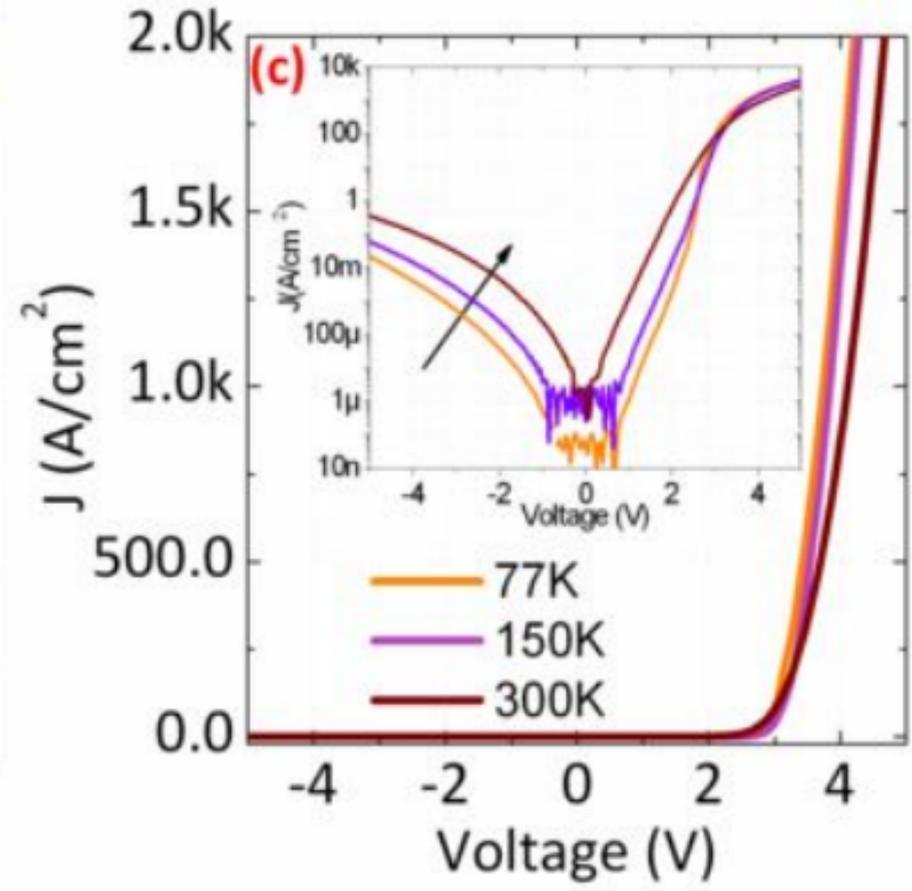

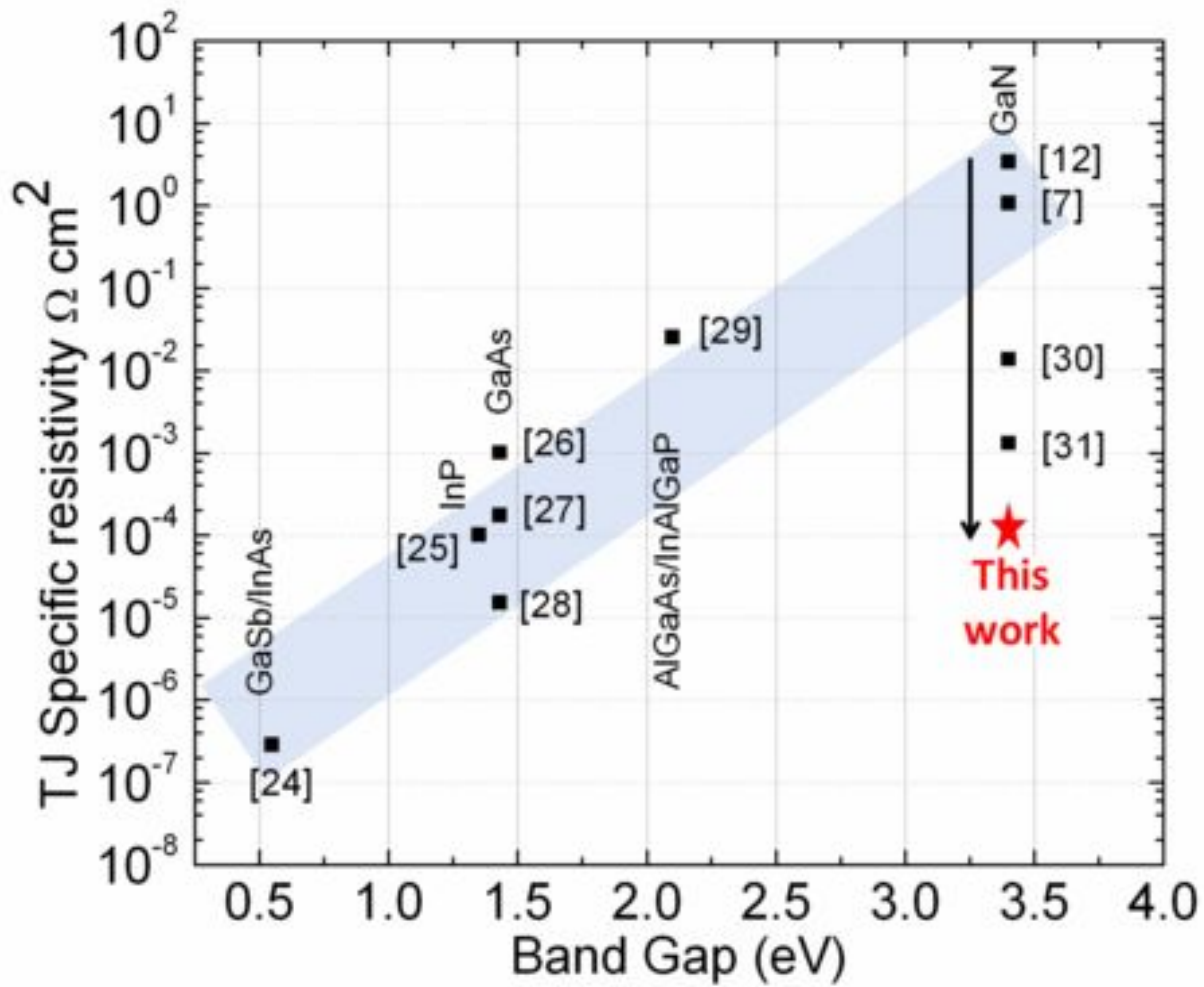